\documentclass[preprint]{aa}       
\usepackage{graphicx}
\usepackage{amssymb}
\usepackage{amsmath}
\usepackage{longtable}
\usepackage{supertabular}
\usepackage{natbib}

\begin{document}

\title{UVMULTIFIT: A versatile tool for fitting astronomical radio interferometric data}

\author{I. Mart\'{\i}-Vidal\inst{1} \and W. H. T. Vlemmings\inst{1} \and S. Muller\inst{1} \and S. Casey\inst{1}
    }

\institute{
European ALMA Regional Center (Nordic Node), Onsala Space Observatory (Chalmers University of Technology), Observatoriev\"agen 90, SE-43992 Onsala, Sweden}

\offprints{\email{mivan@chalmers.se}}

\date{Accepted for publication in A\&A}

\abstract
{The analysis of astronomical interferometric data is often performed on the
images obtained after deconvolving the interferometer's point spread function (PSF). This strategy can be understood (especially for cases of sparse arrays) as fitting models to models, since the deconvolved images are already non-unique model representations of the actual data (i.e., the visibilities). Indeed, the interferometric images may be affected by visibility gridding, weighting schemes (e.g., natural vs. uniform), and the particulars of the (non-linear) deconvolution algorithms. Fitting models to the direct interferometric observables (i.e., the visibilities) is preferable in the cases of simple (analytical) sky intensity distributions.}   
{We present UVMULTIFIT, a versatile library for fitting visibility data, implemented in a Python-based framework. Our software is currently based on the CASA package, but can be easily adapted to other analysis packages, provided they have a Python API.}
{The user can simultaneously fit an indefinite number of source components 
to the data, each of which depend on any algebraic combination of fitting 
parameters. Fits to individual spectral-line 
channels or simultaneous fits to all frequency channels are allowed. }
{We have tested the software with synthetic data and with real observations. In some cases (e.g., sources with sizes smaller than the diffraction limit of the interferometer), the results from the fit to the visibilities (e.g., spectra of close by sources) are far superior to 
the output obtained from the mere analysis of the deconvolved images.}
{UVMULTIFIT is a powerful improvement of existing tasks to extract the maximum amount of information 
from visibility data, especially in cases close to the sensitivity/resolution 
limits of interferometric observations.}


\authorrunning{Mart\'i-Vidal et al.}
\titlerunning{A model-fitting tool for interferometric data.}

\maketitle

\section{Introduction} \label{intro}

An astronomical interferometer does not scan the sky intensity distribution directly; instead, it takes samples of its Fourier transform\footnote{Valid for small fields of view (e.g. \citealt{Cornwell08})} \citep[these are the so-called {\em visibilities}; e.g.,][]{TMS}. If the number of elements in the interferometer is large and the spatial distribution of the array is not sparse, the structure of the observed sources can be recovered with a high fidelity, using non-linear image-reconstruction (i.e., deconvolution) algorithms \citep[e.g.,][]{Cornwell}. However, if the array is sparse (e.g., very long baseline interferometry, VLBI), the image reconstruction may depend strongly on the particulars of the deconvolution algorithm. Hence, the images resulting from these reconstruction algorithms are not {\em direct} and {\em unique} representations of the data, but rather non-unique {\em interpretations} of the actual measurements. Any analysis based on interferometric images (especially those coming from sparse arrays) therefore needs to be understood as {\em fitting models to models}, in the sense that the images themselves are the result of a non-linear mapping from the Fourier domain into the sky plane. This problem can be especially important when comparing images at different frequencies, epochs, or taken with different interferometers, unless the imaging process of the different datasets is performed self-consistently. For instance, \cite{IMV93J} compared a self-consistent image-based analysis and a visibility model-fitting analysis, both applied to real VLBI observations of a radio supernova.

The structure of the observed sources is often simple and/or can be parametrized using simple models (e.g., circular or elliptical rings, discs, or Gaussians; or as a linear combination of simple components). In these cases, it is possible to bypass the imaging of the visibilities and work directly on the interferometric measurements, fitting a model to the visibilities instead of to the images obtained from deconvolution algorithms. This direct approach can be more fruitful than imaging in cases when the sources are very small (compared with the synthesized beam) and/or have a low signal-to-noise ratio. This conclusion is also valid for non-sparse arrays with a large number of antennas such as J-VLA or ALMA.
In particular, it is well-known in astronomical interferometry that an unlimited over-resolution power can be achieved if the dynamic range of the observations is arbitrarily large. Indeed, the smallest resolvable size of a source, $\theta_{min}$, observed with an interferometer, can be written as \citep[e.g.,][]{IMVOver}

\begin{equation}
\theta_{min} = \beta \left(\frac{\lambda_c}{2}\right)^{1/4} \left( \frac{1}{\textrm{S/N}}\right)^{1/2}\times\theta_{beam},
\label{OverResEq}
\end{equation}

\noindent where S/N is the signal-to-noise ratio of the averaged visibilities; $\beta$ weakly depends on the spatial distribution of the telescopes (it tipically takes values between 0.5 and 1); $\theta_{beam}$ is the full width at half maximum (FWHM) of the synthesized beam using natural weighting; and $\lambda_c$ depends on the probability cutoff for a false size-detection (i.e., the chance of measuring a small size for a source that is point-like). The value of $\lambda_c$ is 3.84 for a $2\,\sigma$ cutoff. Equation \ref{OverResEq} assumes that the source size is estimated directly from the visibilities, by means of model fitting. 
Hence, science goals involving observations of simple sources with sizes close to (or smaller than) the resolution limits of the interferometers, will additionally benefit when the imaging process is avoided and the model fitting is performed on the visibilities. Another advantage of model-fitting to the visibilities is that visibilities have uncorrelated noise, whereas the noise of neighbouring pixels in an image is correlated by the effect of the finite size of the point spread function (PSF). Hence, a least-squares fit to the visibilities does not have to deal with a non-diagonal covariance matrix.

Needless to say that there are many cases where the image deconvolution is required 
(e.g., complex structures, such as spiral arms or warped discs), as well as cases where both approaches, imaging and model-fitting, lead to similar results.

In this publication, we present {\sc uvmultifit}, an object-oriented versatile library for model-fitting to the visibilities. To our knowledge, there are currently five main software packages dedicated to the calibration and analysis of astronomical radio-interferometric data. These are the Astronomical Image Processing System ({\sc aips})\footnote{National Radio Astronomy Observatory (NRAO), USA}, {\sc{difmap}} \citep{difmap}, {\sc{miriad}} \citep{miriad}, {\sc{gildas}}\footnote{\texttt{http://www.iram.fr/IRAMFR/GILDAS}}, and the Common Astronomy Software Applications ({\sc casa})\footnote{\texttt{http://casa.nrao.edu}}. All these packages have their own tools for visibility model-fitting, whose main characteristics are summarized in Appendix \ref{OthersApp}. In the same appendix, we compare these packages to {\sc uvmultifit}, and summarize some unique features of the latter.

\section{Description of {\sc uvmultifit}}

The {\sc uvmultifit} interface is written in Python and makes use of \texttt{casapy} (it depends on the {\sc casa} package), although it can be easily adapted to other interferometry software packages (e.g., {\sc aips} via ParselTongue; \citealt{Parsel}). It also uses a C++ extension module for least-squares fitting in a multi-threading environment, which means that parallelization is possible in a multi-core machine.

\subsection{Installation and use}
\label{Install}

We distribute {\sc uvmultifit} under the terms of the general public licence (GPL); the code is open and free. The whole package (together with its documentation) can be downloaded from\\

\noindent \texttt{http://nordic-alma.se/support/software-tools} \\

The instructions for compiling and installing the module can be found in the documentation. After installing it, the user starts {\sc casa} and imports the {\sc uvmultifit} module (for instance under the name \texttt{uvm}). To perform a fit, the user just creates a fit instance, for example,\\

\noindent \texttt{myfit = uvm.uvmultifit(vis="myvis.ms", model=["delta"], var=["0,0,p[0]"], p\_ini=[1.0])}\\

The function \texttt{uvmultifit} depends on several keywords (such as \texttt{vis} or \texttt{model}), whose meaning and syntax are fully described in the software documentation (some basics are given in Sect. \ref{Syntax}). To access the help text, just execute\\

\noindent \texttt{help(uvm.uvmultifit)}.\\

In the example given above, a centred delta component will be fitted to the visibilities of the measurement set called ``myvis.ms''. After the fitting, the final values of the parameters, together with their estimated uncertainties can be recovered by either looking at the \texttt{myfit.result} dictionary or at the ascii file called ``modelfit.dat'' (created in the current working directory).

\subsection{Model-fitting algorithm}

For a given set of interferometric observations, the coordinates of the antenna baselines are taken for each integration time. Then, the Fourier transform of the model of the sky intensity distribution (weighted with a Gaussian approximation of the antenna primary beam) is computed at the $u$ and $v$ coordinates of the baselines. These coordinates are given in units of the observing wavelength. The values of the Fourier transform, computed for each baseline, time, and frequency channel, are taken as the model visibilities, $v^{mod}$, which are then compared to the measured visibilities, $v^{obs}$, using the $\chi^2$ statistic with

$$ \chi^2 = \sum_i{\left|\frac{v^{obs}_i - v^{mod}_i}{\sigma_i}\right|^2}.$$

\noindent In this equation, the index $i$ runs over baselines, times, and frequency channels; $\sigma_i$ is the uncertainty of the $i$-th visibility. The minimization of the $\chi^2$, as a function of the parameters that define the model of the sky intensity distribution, is performed using either the Levenberg-Marquardt approach \citep{Levenberg} or the SIMPLEX algorithm \citep{Nesa}, depending on the choice of the user. The parameter uncertainties are estimated from the post-fit covariance matrix of the parameters (only if the Levenberg-Marquardt algorithm is used), which is scaled by a global factor so that the reduced $\chi^2$ equals its expected value (unity).

The minimization can be performed in two ways. Either the visibilities at all frequency channels are used together in one single fit, or each frequency channel is fitted independently. The user selects between these two approaches by setting the boolean keyword named \texttt{OneFitPerChannel}, which is self-explanatory (see documentation and Sect. \ref{Tests} for examples).

We note that fitting of mosaic data involves phase shifts and baseline re-projections, which are performed by {\sc uvmultifit} before the primary-beam corrections and the Fourier-transforms are applied. This way, {\sc uvmultifit} can be used to fit mosaic data. However, there are two main limitations in our mosaic-fitting algorithm, which restrict the fit to cases when 

\begin{enumerate}
\item all the source components are small compared with the primary beam, and

\item the primary beam is also small, so there is no need to perform holographic projection (i.e., w-term effects) {\em within a pointing}.

\end{enumerate}

In other words, the fit allows us to have different source components spread over a large portion of the sky (there is no limit for the size of the observed sky region), but the size of each individual component must be small compared with the telescope's primary beam (which in turn cannot be large; a few arc-minutes at most). 

If the user needs to fit large sources to the data (sources that cover several antenna pointings, such as ALMA observations of molecular shells in nearby evolved stars), we have developed an extension called {\sc immultifit}. With this extension, the beam-corrected (and gridded) visibilities are computed from both the dirty image and the image of the PSF generated by {\sc casa}, which have all the primary-beam corrections properly applied. A fit with {\sc immultifit} can then be understood as either a fit to the gridded (and beam-corrected) visibilities or as a fit to the dirty image, using models convolved with the PSF. For more information about {\sc immultifit}, we refer to the documentation, accessible by executing \texttt{help(uvm.immultifit)}.

\subsection{Basic syntax}
\label{Syntax}

The model components used to fit the sky intensity distribution are provided in the keyword named \texttt{model} and are specified as a list of strings. For instance, in the example given in Sect. \ref{Install}, the \texttt{model} keyword was set to [``delta''], since the fitting model consisted of only one point source. If the user needs to fit the sum of a point source and a Gaussian source to the data, the \texttt{model} keyword should be set to [``delta'', ``Gaussian'']. 

Each model component used by {\sc uvmultifit} depends on several variables (which must be provided in the keyword \texttt{var}). The variables for a given model component are specified as one single string (the variables are separated by commas within the string). The ordering of the variables is

\begin{enumerate}

\item Right ascension offset (RA, in arcsec).
\item Declination offset (Dec, in arcsec).
\item Total flux density (in Jy). 
\item Major axis (in arcsec).
\item Axis ratio. 
\item Position angle of the major axis (in degrees, from north to east). 

\end{enumerate}

The last three variables are not defined for point sources. For instance, a point source (i.e., a ``delta'') located at an offset of 5.0\,arcsec to the east of the phase centre, and with a flux density of 0.7\,Jy, has its variables defined by the string ``0.5,~0.,~0.7''. 

Each variable of the model component(s) can be set to an arbitrary function of the fitting parameters  
and the observing frequency. That is, {\em the strings that represent the variables can be set to any function written with Python syntax}. For the use of special functions (such as trigonometric or logarithmic functions) the user must call them through \texttt{numpy} (which is internally loaded by {\sc uvmultifit} with the name ``\texttt{np}''). The {\em i-th} parameter is represented by the string ``p[i]'' and the observing frequency (given in Hz) is represented by the string ``nu'' (see the documentation for a more detailed explanation). For instance, the frequency multiplied by the sine of the third fitting parameter is expressed with the string ``nu*np.sin(p[3])''.

As an example for a model component, we created a point-source whose position and flux density need to be determined. In this case, p[0] could be the RA offset, p[1] the Dec offset, and p[2] the flux density. The variables of this delta model component would then be given by the string ``p[0],~p[1],~p[2]''. 
If the source is located at the phase centre and we only need to solve for its flux density, the variables will instead be given by the string ``0,~0,~p[0]'' and, in this case, the only fitting parameter, ``p[0]'', will be the flux density of the source.

\subsection{Examples}

Here we show some illustrative examples of model structures that can be easily implemented in {\sc uvmultifit}. The flexibility of our model-fitting software makes it possible to build quite complicated structures from a superposition of a few simple models. More illustrative examples and details about the syntax of {\sc uvmultifit}, are provided in the software documentation. In the following subsections, we give some useful examples of source models. 

\begin{figure*}[ht!]
\centering
\includegraphics[width=18cm]{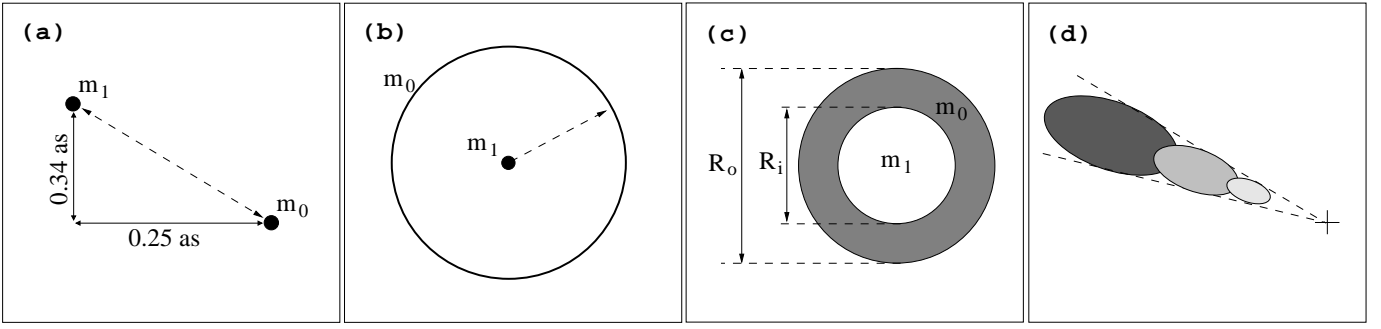}
\caption{Sketches of some of the example models given in the text. $m_i$ refers to the $i$-th model component. (a) are the two point sources with known (i.e., {\em fixed}) separation, although the absolute separation of the compound source can be unknown; (b) is the ring with a point source at the centre; (c) is a disc with a hole; and (d) is a jet with synchrotron self-absorption (the ellipses represent the FWHM of the fitting Gaussians, darker-grey tones represent lower frequencies, and the cross marks the location of the jet base).}
\label{Figmods}
\end{figure*}

\subsubsection{Point source with unknown position and spectral index}

In this example, we only have one model component of type ``delta''. Hence,

\begin{itemize}
\item \texttt{model} = [``delta'']
\item \texttt{var} = [``p[0],~p[1],~p[2]*(nu/1.e9)**p[3]''].
\end{itemize}

The parameter p[2] is the flux density at 1\,GHz and p[3] is the spectral index. The other two parameters, p[0] and p[1], are the RA offset and Dec offset, respectively.

\subsubsection{Two point sources with known separation}

For two point sources separated by 0.25\,arcsec and 0.34\,arcsec (in RA and Dec, respectively) one from the other, but with unknown absolute positions and flux densities, the corresponding model would be

\begin{itemize}
\item \texttt{model} = [``delta'',~``delta'']
\item \texttt{var[0]} = ``p[0],~p[1],~p[2]'' 
\item \texttt{var[1]} = ``p[0]+0.25,~p[1]+0.34,~p[3]''.
\end{itemize}

Note that \texttt{model} is a list with the type of model components, that is, two point sources (deltas);  \texttt{var[0]} and \texttt{var[1]} are the two strings that define the variables of the two components (i.e., \texttt{var} is a list of two strings). Each delta component has three variables (RA offset, Dec offset, and flux density). In regard to the fitting parameters, p[0] and p[1] are the RA and Dec offsets of the first delta, p[2] is its flux density, and p[3] is the flux density of the second delta. A sketch of this model is shown in Fig.\,\ref{Figmods}(a).

\subsubsection{Narrow ring with a delta at its centre}

The absolute position of the compound source is unknown (i.e., we fit it) and the infinitely narrow ring (of unknown size) is assumed to be circular. The corresponding model is

\begin{itemize}
\item \texttt{model} = [``ring'',~``delta'']
\item \texttt{var[0]} = ``p[0],~p[1],~p[2],~p[3],~1,~0''
\item \texttt{var[1]} = ``p[0],~p[1],~p[4]*(nu/1.e9)**p[5]''.

\end{itemize}

In this case, p[0] and p[1] are the RA and Dec offsets of the two components (delta and ring); p[2] is the total flux density of the ring and p[3] is its diameter; p[4] is the flux density of the delta at 1\,GHz; and p[5] is the spectral index of the delta. Note that the axis ratio of the ring is set to unity and the position angle is set to zero (although any value of the position angle would work in this case). 
A sketch of this model is shown in Fig.\,\ref{Figmods}(b).

\subsubsection{Disc with a hole at its centre}
\label{DiskWithHole}

A circularly symmetric disc with a hole at its centre can be built as the addition of two discs of different size; one disc with a larger diameter, $R_o$, and positive flux density, $F_o$, and another disc with a smaller diameter, $R_i$, and negative flux density, $-F_i$. The total flux density of the disc with the hole is then $F_o - F_i$. If the emission intensity is exactly zero in the region of the hole, the flux density per unit beam (i.e. the intensity) of both discs must be equal in absolute value, so that the effect of the inner negative disc will be the exact subtraction of the flux density from the inner side of the larger disc. Hence, $F_i$ is related to both $F_o$ and the size ratio, $a = R_i/R_o$, in the way 

$$ F_i = a^2 F_o.$$

We now define p[0] as the overall flux-density of the disc with hole, $F_o - F_i$; p[1] as the outer size of the disc, $R_o$; and p[2] as the relative disc thickness, $a$. With these definitions, we finally obtain

\begin{itemize}
\item \texttt{model} = [``disc'',~``disc'']
\item \texttt{var[0]} = ``0,~0,~p[0]*(1+p[2]**2),~p[1],~1,~0''
\item \texttt{var[1]} = ``0,~0,~$-$p[0]*p[2]**2,~p[1]*p[2],~1,~0''.
\end{itemize}

Needless to say that if any of these parameters ($F_o-F_i$, $R_o$, and/or $a$) is known, we can fix it to its exact value (hence decreasing the number of fitting parameters). We can also build an elliptical disc with a hole (and even solve for its ellipticity) by changing the fifth (and sixth) variables of both discs equally.
A sketch of this model is shown in Fig. \ref{Figmods}(c).

\subsubsection{Compact AGN jet with a core-shift}
\label{AGNmodel}

In a radio-loud active galactic nucleus (AGN) whose jet can be resolved with VLBI, the location of the peak intensity in the jet (i.e., the jet core) depends on the observing frequency. This is the so-called {\em core-shift} effect and is produced by synchrotron self-absorption (SSA) at the region of the jet close to its base \citep[e.g.][]{Lobanov}. Basically, SSA is higher at shorter distances to the jet base. Since SSA decreases with increasing frequency, the peak intensity at higher frequencies will be closer to the jet base.
In the case of broadband VLBI observations of the jet cores in AGN, it may be possible to observe the core-shift effect throughout the observed band. We model the jet core (at a given frequency) using an elliptical Gaussian. The main axis of the Gaussian will be oriented in the direction of the jet, which is indeed the direction of the core-shift effect. Since the cores at frequencies $\nu_1$ and $\nu_2$ are separated by \citep{Lobanov} 

\begin{equation}
\Delta r = \Omega \left(\frac{1}{\nu_2} - \frac{1}{\nu_1}\right),
\label{LobEq}
\end{equation}

\noindent we can model the core of an AGN jet by defining the Gaussian variables in the following way: \\

\noindent RA $\rightarrow$  p[0] + p[6]*(1/nu)*np.sin(p[5]) \\
Dec $\rightarrow$      p[1] + p[6]*(1/nu)*np.cos(p[5]) \\
Flux density $\rightarrow$     p[2]*(nu/1.e9)**p[7]) \\
Diameter $\rightarrow$         p[3]/(nu/1.e9)**p[8]) \\
Axis ratio $\rightarrow$            p[4] \\
Pos. angle $\rightarrow$       p[5]*180./np.pi .\\

On the one hand, the offsets in right ascension and declination of the jet base (i.e., the core at $\nu\rightarrow\infty$) are p[0] and p[1], respectively; the flux density at 1\,GHz is p[2], the jet diameter at 1\,GHz is p[3], the axis ratio of the core Gaussian (assumed to be equal at all frequencies) is p[4], and the position angle of the jet (in radians and at all frequencies) is p[5]. On the other hand, p[6] is the normalized core-shift (i.e., $\Omega$ in Eq. \ref{LobEq}), p[7] is the spectral index of the core, and p[8] would be related to the jet shape (it should be set to unity for a conical jet, which is a reasonable assumption at cm wavelengths). 
A sketch of this model is shown in Fig. \ref{Figmods}(d).
We note that this model can also be used to simultaneously fit a set of (phase-referenced) narrow-band VLBI observations of an AGN, taken at different frequencies (but close by in time). In this case, the \texttt{vis} keyword could be set either to a list of strings (i.e., the names of all the measurement sets; one set for each frequency band) or to one string (i.e., the name of a measurement set resulting from the concatenation of the data at all frequencies).

\section{Tests with synthetic and real data}
\label{Tests}

In this section, we present some results of {\sc uvmultifit} applied to both synthetic and real data. On the one hand, we use the simulated data to compare the best-fit model parameters with those used in the generation of the synthetic visibilities. On the other hand, we use real data for which visibility model-fitting results have previously been reported, so that we can compare the results of {\sc uvmultifit} with those found in the literature.

\subsection{Synthetic data}

\subsubsection{Source model}
\label{SourceSec}

The simulated source model consists of two point-like emitters separated by 1\,arcsec. The declination is set to $-30$\,deg. We have generated fake spectra for both sources, consisting of random absorption/emission lines, defined between 98\,GHz and 102\,GHz, with random strengths. The spectrum of one of the sources is set to be equal to that of the other source, but with a small frequency shift. The position of the weakest source (20\,mJy) is set at the phase centre of the image and the second source (29\,mJy) is shifted 1\,arcsec to the west. The spectra of both sources are shown in Fig. \ref{FitTwoDeltas} (top).

\subsubsection{Simulated interferometer}
\label{ArraySec}

To generate the synthetic visibilities, we used the {\sc casa} task \texttt{simobserve}. This task allows us to select different array configurations for many interferometers (e.g., the sub-milimeter array, SMA, or the Atacama large mm/submm array, ALMA) and generate synthetic data for any model, using realistic noise conditions for the antennas and the atmosphere. In our simulations with \texttt{simobserve}, we made use of the ALMA interferometer in its compact configuration at cycle 0. This configuration consists of 16 antennas separated by a maximum baseline of $\sim$120\,m. We added realistic noise to the data (the typical noise for the ALMA antennas at 100\,GHz, according to the database used by \texttt{simobserve}). This noise accounts for the system temperatures of the antennas and for the atmospheric opacity. The length of the simulated observations was set to 2\,hours, centred on the source transit.

\subsubsection{Results from imaging}
\label{Results}

We note that the source separation is 4--5 times {\em smaller} than the FWHM of the restoring beam from the simulated ALMA cycle 0 compact configuration (4.68$\times$4.15\,arcsec, using {\em Briggs} weighting with a robustness parameter of 0.5). Hence, the CLEAN image of this source is obviously point-like. The spectrum of the single source detected in the image is the combination of the spectra from the two close-by components used in the simulation. The true spectra of the two components are shown in Fig. \ref{FitTwoDeltas}(a) and the CLEANed spectrum is shown in Fig. \ref{FitTwoDeltas}(b). As can be seen in the figure, the two spectra in the image plane cannot be separated. The dynamic range achieved in the CLEANed image cube (peak line over rms) is $\sim50$.

\subsubsection{Visibility model-fitting. Locating the sources}
\label{SpecDeltas}

The fit to the visibilities is performed in two steps. In the first step, we select a subset of frequency channels corresponding to continuum emission and fit the data with \texttt{OneFitPerChannel}=False (i.e., one single fit to all the frequency channels; see the documentation for more details). The line-free channels are selected from visual inspection of the source spectrum (Fig. \ref{FitTwoDeltas}b). In this fit to the continuum emission, the position of the strongest component is left free, with no bounds, whereas the {\em relative} position of the second component to the first one is bounded in RA and Dec to be within 1\,arcsec from its true value. The initial guesses of the relative coordinates are set at 0.1 and 0.2\,arcsec from the true values of RA and Dec, respectively. We note that in similar cases where the relative position of two sources is well known (e.g., from observations with other instruments and/or at other frequencies; see Sect. \ref{RealData} for a real example), the strategy of parametrizing the positions of the two components as the absolute position of the first component plus the relative position of the second one maximizes the amount of a-priori information used in the fit. This strategy is {\em not possible} if the analysis is performed on the image plane using deconvolution algorithms.

\subsubsection{Visibility model-fitting. Recovering the spectra}

\begin{figure*}[ht!]
\centering
\includegraphics[width=15cm]{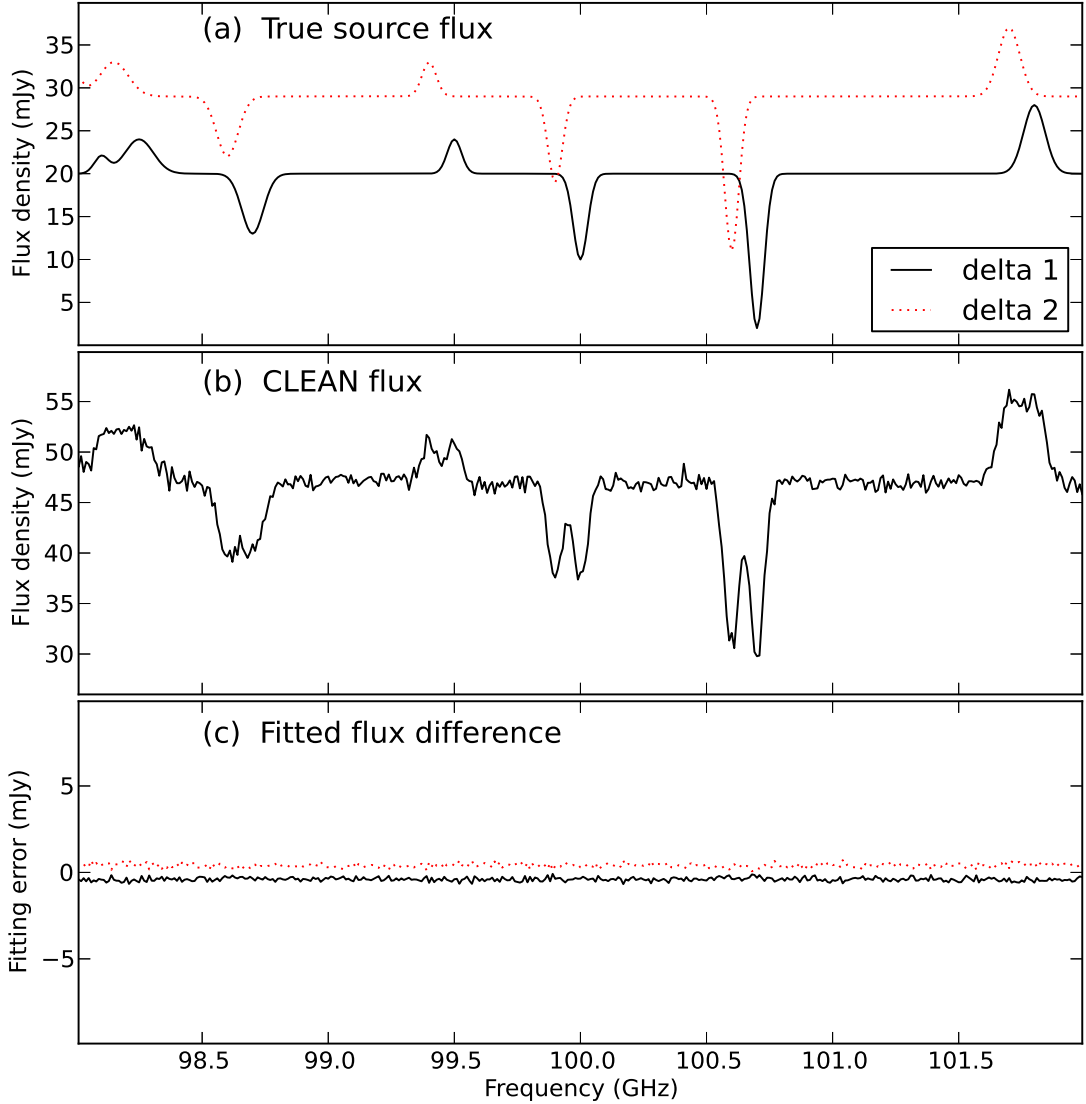}
\caption{Top: model flux densities used in our simulation of two close-by point sources. Delta 1 is a point source located at the phase centre and delta 2 is a point source located 1\,arcsec to the west of delta 1. Middle: spectrum of the point-like source obtained from the image deconvolution. Bottom: difference between the simulated flux densities and those fitted with {\sc uvmultifit}.}
\label{FitTwoDeltas}
\end{figure*}

The positions estimated from the fit described in the previous section are then {\em fixed} in a second fit, where we only solve for the fluxes of the two delta components at each spectral channel (i.e., we now set \texttt{OneFitPerChannel}=True). 
The residuals from this second fit (i.e., the difference between the true flux densities and the fitted flux densities) are shown in Fig. \ref{FitTwoDeltas}(c). We call these quantities {\em fitting errors}).
We note that the positions fitted in continuum mode only differ by about 15\,mas (at most) from the true positions of the sources. In addition, the spectrum of one source is completely decoupled from that of the other, since there is no line emission/absorption features in the fitting errors, as shown in Fig. \ref{FitTwoDeltas}(c). The typical fitting error is around 0.5\,mJy for both components. We note that the flux-density fitting errors for the strongest source are always positive, while the errors for the weakest source are negative. These systematics (which are about 2\% of the flux density) are directly related to the small difference between the real and the fitted positions of the sources. These small shifts map into biases in the absolute fluxes of the individual components, although the frequency dependence of these fluxes is flat (so the shapes of the source spectra are not affected).

\subsection{Real data}
\label{RealData}

We used {\sc uvmultifit} in the analysis of real data where the science goals were critically affected by the limited spatial resolution. As real-case examples, we show here the results obtained from the analysis of ALMA cycle-0 data on the lensed blazar PKS\,1830$-$211\footnote{Project ADS/JAO.ALMA\#2011.0.00405.S} and VLBI data of the radio supernova SN\,1993J.

\subsubsection{ALMA observations of PKS\,1830$-$211}

The separation between the two lensed images of the blazar (one at the north-east, NE, and the other at the south-
west, SW, of the foreground galaxy) is $\sim$1\,arcsec. However, the restoring beam of the ALMA B3 ($\sim$90\,GHz) observations is 1.5$\times$1.9\,arcsec, with a position angle of 80 deg. The two images are thus not resolved by the synthesized beam.

Molecular absorption from the foreground lensing galaxy is found along both lines of sight to the SW and NE images, although with a slight offset in velocity (due to the galactic rotation, \citealt{Muller}).
Hence, this source is qualitatively similar to the simulated source described in Sect. \ref{SpecDeltas}. Indeed, we used the same fitting strategy to extract the absorption spectra toward the NE and SW images.

\begin{figure}[ht!]
\centering
\includegraphics[width=9cm]{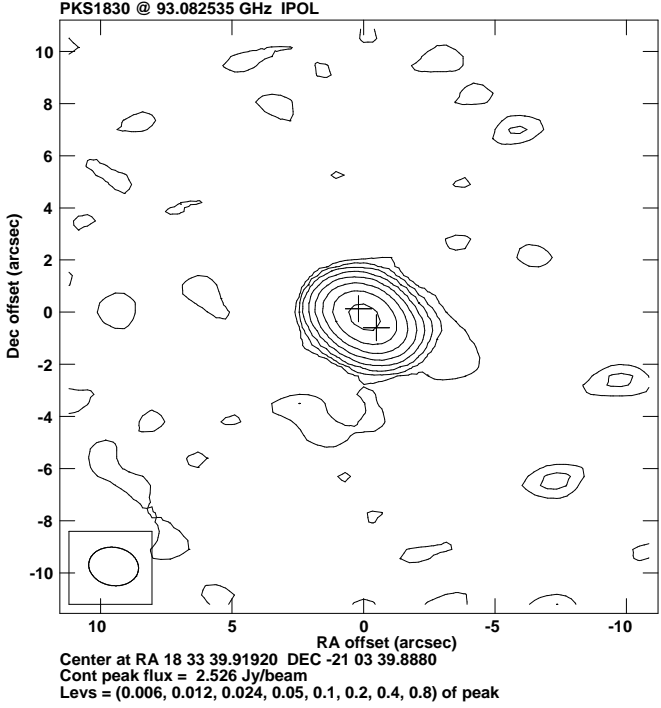}
\caption{ALMA clean image of PKS\,1830$-$211 at 93\,GHz (only one spectral channel, of 977\,kHz width, has been used). The FWHM of the restoring beam is shown at bottom left. The best-fit position estimates for the NE and SW images are marked with crosses.}
\label{PKSImage}
\end{figure}

\begin{figure}[ht!]
\centering
\includegraphics[width=9cm]{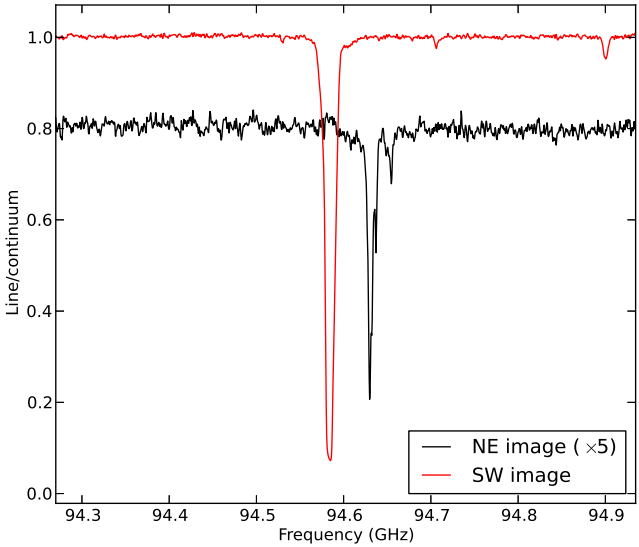}
\caption{Normalized spectra of the NE and SW images of PKS\,1830$-$211, obtained with {\sc uvmultifit}. For clarity reasons, the amplitude of the NE spectrum has been multiplied by 5 (the absorption in the NE image is much weaker than in the SW image) and its continuum has been normalized to 0.8.}
\label{PKSFig}
\end{figure}

The image of PKS\,1830$-$211, obtained using only one spectral channel, is shown in Fig. \ref{PKSImage}. The location of the NE and SW images (as estimated with {\sc uvmultifit}) is marked with crosses. 
We used {\sc uvmultifit} to estimate the source positions from the continuum parts of the spectra. Since the relative position between the NE and SW images is well-known from VLBI observations (e.g., \citealt{Jin}), we bounded the fit of the relative positions to be within only 10\,mas with respect to the VLBI-astrometry results. After estimating the positions of the two point sources, we fixed them and recovered the spectra of each point source by fitting the flux densities for each spectral channel (i.e., the flux densities are the only fitting parameters in the last fit). We show a fraction of the resulting spectrum in Fig. \ref{PKSFig}. The spectra from both the NE and the SW image are completely uncorrelated.  
We note that if larger deviations (of even 0.1\,arcsec) are allowed in the fit of the relative position between the point sources, the final results are basically unchanged; the small changes (of $\sim$2\%) in the best-fit position offsets of the sources map into very small changes (of $\sim$0.5\%) in the fitted fluxes, but the normalized absorption spectra (Fig. \ref{PKSFig}) remain unchanged.
More details about these results will be given elsewhere (Muller at al. in prep.).

\subsubsection{VLBI observations of SN\,1993J}
\label{VLBIData}

Supernova SN\,1993J was observed with VLBI at different frequencies for more than one decade (see \citealt{IMV93J} and references therein for details), using the AGN in M\,81 (hereafter M\,81*) as the phase calibrator. The position of the supernova was used as a phase-reference to analyze the frequency dependence of the structure in M\,81* jet, as well as its time evolution (\citealt{IMVM81}, and references therein).

Here we show an example of a fit to the visibilities of SN\,1993J, combining in one single fit data taken close by in time at three different frequencies. We selected VLBI observations of SN\,1993J at 1.7, 5.0, and 8.4\,GHz, taken on 15 Nov 1997.
The structure of the supernova can be well described as a spherical shell with a thickness of $\sim$30\% of its radius (\citealt{IMV93J}). We can construct this spherical-shell model as the addition of two concentric filled spheres; a larger one with a positive flux density and a smaller one with a negative flux density. This approach is similar to that used for the disc with a hole described in Sect. \ref{DiskWithHole}.

We used the same spherical-shell model at all the three frequencies. However, the core of the M\,81* jet at different frequencies is located at slightly different positions on the sky, due to synchrotron self-absorption (this is the basic idea behind the AGN jet model described in Sect. \ref{AGNmodel}). Hence, the relative position between M\,81* and SN\,1993J depends on the observing frequency. We implemented this case in a way similar to that described in Sect. \ref{AGNmodel}.
On the one hand, the larger sphere is described with the following variables:\\

\begin{itemize}
\item RA = ``p[0] + 1.e9*p[2]*np.sin(p[3]*np.pi/180.)/nu''
\item Dec = ``p[1] + 1.e9*p[2]*np.cos(p[3]*np.pi/180.)/nu''
\item Flux = ``p[4]*(nu/2.3e9)**p[5]''
\item Size = ``p[6]''. 
\end{itemize}

On the other hand, the RA and Dec offsets of the smaller sphere are set to be equal to those of the larger sphere, and the flux and size of the smaller sphere are set to

\begin{itemize}
\item Flux = ``$-$(0.7**3.)*p[4]*(nu/2.3e9)**p[5]''
\item Size = ``0.7*p[6]''.
\end{itemize}
 
Here, the ratio of the inner shell radius to the outer shell radius is set to 0.7 (\citealt{IMV93J}). Parameters p[0] and p[1] are the RA and Dec offsets of the supernova shell centre, referred to as the jet base of M\,81* (see Eq. \ref{LobEq}); parameter p[2] is the normalized core-shift of M\,81* (i.e., $\Omega$ in Eq. \ref{LobEq}) in units of arcsec\,GHz$^{-1}$; parameter p[3] is the position angle of the core shift in M\,81* (which should be equal to the direction of the jet, if we do not consider the precessing motion reported in \citealt{IMVM81}); parameter p[4] is proportional to the flux density at 2.3\,GHz, and p[5] is the spectral index of the shell; finally, p[6] is the size of the shell, assumed to be equal at all frequencies.

The simultaneous fit to the three datasets gives us a core shift of $2.05\pm0.13$\,mas\,GHz$^ {-1}$ (the fit was bounded between 0 and 10\,mas\,GHz$^ {-1}$). This value is consistent with the core shift reported in \cite{IMVM81} from the analysis of all the available VLBI epochs ($1.75\pm0.20$\,mas\,GHz$^ {-1}$). In addition, our estimate of the position angle of the core shift is $62\pm4$\,deg. (the fit was bounded between 0 and 180 degrees), compatible with the direction of the M\,81* jet (which is precessing between 60 and 70 degrees). 
Finally, our estimate of the shell radius is $3.23\pm0.02$\,mas, which is also compatible with the values reported in \cite{IMV93J} for the same epochs (i.e., model-fitting average of $3.00\pm0.15$\,mas for the different frequencies\footnote{\cite{IMV93J} used a model of a shell with absorption by the ejecta. They noted that the absorption by the ejecta maps into slightly smaller size estimates.}), 
and the estimated spectral index is $-0.83\pm0.01$, also in agreement with previous reports. The simultaneous fit of the data at the three different frequencies, made with {\sc uvmultifit}, improves the precision of the estimated quantities in all cases. However, this strategy can be applied as long as the structure of the supernova shell is similar at all frequencies (this condition does not hold at later epochs, \citealt{IMV93J}).

This fit may be still further complicated by also adding the visibilities of M\,81*. We could then describe M\,81* as a new model component: a frequency-dependent Gaussian, using the model described in Sect. \ref{AGNmodel}. Then, the major axis of this Gaussian would be set to be aligned (by means of the core-shift effect) to the frequency-dependent position shifts of the supernova shell.

\section{Some tips and tricks}

Here we show a list of some tips for a better use of {\sc uvmultifit}. For a more in-depth discussion, please read the documentation.

\begin{itemize}

\item We have developed a graphical user interface (GUI), based in Qt/PySide\footnote{Distributed under the LGPL license. We provide a pre-compiled version for Linux (64 bit). For other platforms, the user should install the PySide module by him/herself (install it for Python 2.6, which is the version used by {\sc casa} up to version 4.1).}, where the user can set all the properties and parameters of the fit. The GUI also plots the values of the fitted parameters as a function of frequency (when the fit is made with \texttt{OneFitPerChannel}=True) and shows their values after the fit (when the fit is performed with \texttt{OneFitPerChannel}=False). The user can execute the GUI by running the function \texttt{GUI()} of the package. We note, though, that the GUI is currently in beta version (tested on {\sc casa} version 4.1). The code of the GUI can be downloaded from the same location as the code of the main module.

\item If the dataset is large and there are many spectral channels, it may be a good idea to average the data in time (this can be done on-the-fly; see the documentation), as long as the time-smearing effects are small. The user could also restrict the fit to only the spectral channels of interest and/or reduce the spectral resolution (i.e., increase the \texttt{chanwidth} parameter) if possible.

\item There may be cases where the user needs to combine datasets at different frequencies, but the relative astrometry among the different datasets is not very precise. In these cases, the user can define position offsets with fitting parameters that are {\em turned on} just for a set of frequencies. For instance, a RA offset defined as \\ 

``p[0] + p[1]*(np.abs(nu - 1.e9)$<$1.e8)'' \\ 

will fit the offset as p[0] + p[1] for frequencies between 0.9 and 1.1\,GHz, and just as p[0] for the rest of frequencies. More complex logical expressions can also be provided.

\item FITS files or {\sc casa} model images (e.g., the \texttt{*.model} directories generated after running CLEAN) can be used as fixed models in the fitting; the user can either fix these models completely (so they will be effectively subtracted from the data before the fit), or fit their overall flux density to any function of the frequency and/or the fitting parameters. To read an image as a set of delta components, the user can call the function \texttt{modelFromClean()}. 

\item We must note that there are no kinematics (i.e., line profiles) implemented in any of the models listed in Table~\ref{Components}. This is a common limitation of all the visibility-fitting packages. However, {\sc uvmultifit} allows us to overcome this limitation in some cases, since it is possible to define arbitrary frequency-dependent model structures. For instance, an expanding bubble with line emission can be modelled as a ring with a frequency-dependent size (in this case, the size would depend quadratically on the observing frequency).

\end{itemize}

\paragraph{Advanced use of {\sc uvmultifit}} ~\\

In addition to the best-fit values of the fitting parameters, {\sc uvmultifit} also returns a \texttt{uvmultifit} object, whose properties (e.g., model components, variables, and/or fitting parameters) can be changed by the user. This object has a \texttt{fit()} function, which allows the user to re-fit the data several times, without the need of re-loading and re-averaging the dataset each time. This will be specially useful when working with large datasets to check, e.g., for variability at different time scales and/or to compare the fitting quality of different models in an efficient way. In the software documentation, we give an example of the use of {\sc uvmultifit} for the study of time variability during the extent of an interferometric observation.

\section{Summary}
\label{Summary}

We have presented {\sc uvmultifit}, an object-oriented combination of Python and C++ modules that allow the user to fit astronomical interferometric data as an arbitrary combination of different model components. The user can define any algebraic relationship among the variables that define the different components, so that relatively complex structures can be parametrized from a wise combination of simple source intensity distributions. 

The fits can be performed in two main modes, namely, to each independent frequency channel or to all frequencies at once. In any of these cases, the variables that define the models can explicitly depend on the frequency in an arbitrary way. 

We showed some examples of complex source structures that can be parametrized with {\sc uvmultifit}, and checked the code against simulated and real observations. 
The object-oriented structure of {\sc uvmultifit}, combined with the run-time compilation of the models and the powerful scripting capabilities of {\sc casa} (and eventually ParselTongue, if {\sc uvmultifit} is adapted for its use in {\sc aips}) make {\sc uvmultifit} a powerful tool for an advanced model-fitting of astronomical interferometric observations. Using this code to determine time variability at different time scales, compare different fitting models of the same dataset, and/or define complex source structures as the addition of different model components, with complex algebraic relationships among them, will be easy, fast, and efficient.

\begin{acknowledgements}
The National Radio Astronomy Observatory is a facility of the National Science Foundation
operated under cooperative agreement by Associated Universities, Inc.
\end{acknowledgements}

\bibliographystyle{aa}

\begin{thebibliography}{}

\bibitem[Cornwell \& Willkinson(1981)]{Cornwell} Cornwell, T. J. \& Wilkinson, P. N. 1981, MNRAS, 196, 106

\bibitem[Cornwell et al.(2008)]{Cornwell08} Cornwell, T.~J., Golap, K., \& Bhatnagar, S.\ 2008, IEEE Journal of Selected Topics in Signal Processing, 2, 647 

\bibitem[Kettenis et al.(2006)]{Parsel} Kettenis, M., van Langevelde, H. J., Reynolds, C., \& Cotton, B. 2006, ASPC, 351, 497

\bibitem[Jin et al.(2003)]{Jin} Jin, C., Garrett, M. A., Nair, S., et al. 2003, MNRAS, 340, 1309

\bibitem[Kovalev et al.(2008)]{Kovalev} Kovalev, Y. Y., Lobanov, A. P., Pushkarev, A. B., \& Zensus, J. A. 2008, A\&A,
483, 759

\bibitem[Levenberg(1944)]{Levenberg} K. Levenberg 1944, Quarterly of Applied Mathematics, 2, 164

\bibitem[Lobanov(1998)]{Lobanov} Lobanov, A. P. 1998, A\&A, 330, 79

\bibitem[Mart{\'{\i}}-Vidal et al.(2011a)]{IMV93J} Mart{\'{\i}}-Vidal, I., Marcaide, J.~M., Alberdi, A., et al.\ 2011a, \aap, 526, A142

\bibitem[Mart{\'{\i}}-Vidal et al.(2011b)]{IMVM81} Mart{\'{\i}}-Vidal, I., Marcaide, J.~M., Alberdi, A., et al.\ 2011b, \aap, 533, A111

\bibitem[Mart\'i-Vidal, P\'erez-Torres \& Lobanov(2012)]{IMVOver} Mart\'i-Vidal, I., P\'erez-Torres, M.A., Lobanov, A. 2012, A\&A 541, A135

\bibitem[Muller et al.(2006)]{Muller} Muller, S., Gu{\'e}lin, M., Dumke, M., Lucas, R., \& Combes, F.\ 2006, \aap, 458, 417

\bibitem[Nelder \& Mead(1965)]{Nesa} Nelder, J.~A., Mead, R. 1981, Computer Journal, 7, 308

\bibitem[Sault, Teuben \& Wright(1995)]{miriad} Sault, R.~J., Teuben, 
P.~J., \& Wright, M.~C.~H.\ 1995, Astronomical Data Analysis Software and Systems IV, 77, 433 

\bibitem[Shepherd, Pearson \& Taylor(1994)]{difmap} Shepherd, M.~C., 
Pearson, T.~J., \& Taylor, G.~B.\ 1994, \baas, 26, 987 

\bibitem[Thomson, Moran \& Swenson(1986)]{TMS} Thomson, A. R., Moran, J. M., \& Swenson, G. W. 1986, Interferometry and Synthesis in Radio Astronomy (New York: Wiley)



\end{thebibliography}

\appendix 

\section{Short summary of current visibility model-fitting tools}
\label{OthersApp}

The main software packages currently used to calibrate and analyse standard astronomical radio-interferometric data are {\sc aips}, {\sc difmap}, {\sc miriad}, {\sc gildas}, and {\sc casa}. All these packages have their own tools for visibility model-fitting, whose main characteristics are summarized in Table \ref{Others}.

\begin{table*}
\caption{Currently available interferometry model-fitting tools (Y = YES; N = NO)} 
\label{Others} 
\centering 
\begin{tabular}{l | l l l l l} 
\hline\hline 
Package & Wide-band & \# comp. & Spectral index  & Mosaic & Generic vars. \\ 
\hline 
{\sc casa} (uvmodelfit) &  Y  &  1         &  N        &  N &  N \\
{\sc aips} (uvfit/slime)  &  N  &  60        &  N        &  N &  N \\
{\sc difmap} (modelfit) &  Y &  $\infty^{1}$  &  Y       &  N &  N \\
{\sc miriad} (uvfit)    &  N   &  $\infty$  &  Y$^{2}$ &  N &  N \\
{\sc gildas} (uvfit)    &  N   &  4         &  N        &  N &  N \\
{\sc uvmultifit}        &  Y  &  $\infty$  &  Y$^{3}$ & Y & Y$^{3}$ \\
\hline 
\end{tabular}
\tablefoot{{\em Wide-band} stands for baseline re-projection of each frequency channel to avoid bandwidth smearing; {\em \# comp.} is the largest allowed number of simultaneous components in the fit; {\em spectral index} stands for the possibility of fitting spectral indices to the components; {\em mosaic} stands for the posibility of dealing with observations of a given source with different phase centres (i.e., pointings); and {\em generic vars.} stands for the possibility of using generic functions to tie the variables among the different components.\\
$^{1}$ The ellipticity of all the fitted extended components must coincide.\\
$^{2}$ The spectral index is not fitted, but estimated from the post-fit model parameters.\\
$^{3}$ A generic frequency dependence is allowed.}
\end{table*}

\begin{table*}
\caption{Available model components (Y = YES; N = NO)} 
\label{Components} 
\centering 
\begin{tabular}{l | l l l l l l l l l} 
\hline\hline 
Package & delta & disc & Gaussian  & ring & bubble & sphere & expo & power$-2$ & power$-3$ \\
\hline 
{\sc casa} (uvmodelfit)  & Y & Y & Y & N & N      & N      & N    & N    & N    \\
{\sc aips} (uvfit/slime) & Y & N & Y & N & N      & Y$^{1}$& N    & N    & N    \\
{\sc difmap} (modelfit)  & Y & Y & Y & Y & N      & Y      & N    & N    & N    \\
{\sc miriad} (uvfit)     & Y & Y & Y & Y & Y$^{1}$& Y      & N    & N    & N    \\
{\sc gildas} (uvfit)     & Y & Y & Y & Y & Y      & N      & Y$^1$& Y$^1$& Y$^1$\\
{\sc uvmultifit}         & Y & Y & Y & Y & Y      & Y      & Y    & Y    & Y    \\
\hline 
\end{tabular}
\tablefoot{{\em delta} refers to a point source; {\em disc} to a uniformly bright disk; {\em ring} to an infinitely narrow ring; {\em sphere} refers to an optically thin uniform filled sphere; {\em bubble} to a uniform spherical surface; {\em expo} refers to an exponential radial decrease; {\em power$-2$} refers to a radial, $r$, decrease like $\propto (r^2+r_0^2)^{-1}$ (in this case, the fitted flux is the integral from $r=0$ to $r=r_0$); and {\em power$-3$} refers to a radial decrease like $\propto \left(1+(2^{2/3}-1)\,(r/r_0)^2\right)^{-3/2}$\\
$^{1}$ No ellipticity is allowed.}
\end{table*}

The model components (i.e., sky intensity distributions) that can be used in these packages are listed in Table \ref{Components}. The most commonly used components in all packages are the point source, the Gaussian, the disc, and the uniform (optically thin) filled sphere. Other sky intensity distributions such as rings, bubbles, or an exponential radial decrease are less commonly offered. {\sc uvmultifit} implements all the models listed in Table \ref{Components}. We note that the main advantage of {\sc uvmultifit}, compared with the other packages, is the possibility of handling generic algebraic relationships among the variables that define the different model components. Generic functions of the observing frequency can also be used. In Sect. \ref{Results}, we show some examples of this feature. {\sc uvmultifit} also gives the possibility of working with mosaic data, corrects for primary-beam effects\footnote{This is currently limited to homogeneous arrays, and the approximation of the beam shape is computed from the antenna diameters.}, and minimizes the effect of bandwidth smearing by re-projecting the baselines in Fourier space for each spectral channel; this is especially useful for wide-band interferometric observations (where the bandwidth is a considerable fraction, 5\% or more, of the observing frequency).

\end{document}